\documentstyle[preprint,aps,epsf]{revtex}
\def\lessim{\mathrel {\vcenter {\baselineskip 0pt \kern 0pt
\hbox{$<$} \kern 0pt \hbox{$\sim$} }}}
\def\gessim{\mathrel {\vcenter {\baselineskip 0pt \kern 0pt
\hbox{$>$} \kern 0pt \hbox{$\sim$} }}}
\def \rightdownarrow
 {\kern.4em {\raise1.75ex\hbox{$\Bigl |$}$\kern-0.27em{\longrightarrow}$}}
\def \mrightdownarrow
 {\kern.4em {\raise1.25ex \hbox{$|$} \kern-0.27em{
                                                \longrightarrow}}}

\def\Lambdab{$\overline{\Lambda}^0_b$}
\def\pbarp{$p\bar{p}$}

\def\mMEt{\not\kern-.35em {E_T}}

\begin{document}
\title{
%
%
%
Measurement of $b$~Quark Fragmentation Fractions\\
in \pbarp\ Collisions at $\sqrt{s}=1.8$~TeV
}
\author{
\font\eightit=cmti8
\def\r#1{\ignorespaces $^{#1}$}
\hfilneg
\begin{sloppypar}
\noindent
T.~Affolder,\r {21} H.~Akimoto,\r {42}
A.~Akopian,\r {35} M.~G.~Albrow,\r {10} P.~Amaral,\r 7 S.~R.~Amendolia,\r {31} 
D.~Amidei,\r {24} J.~Antos,\r 1 
G.~Apollinari,\r {35} T.~Arisawa,\r {42}
% ...
\end{sloppypar}
\vskip .026in
\begin{center}
(CDF Collaboration)
\end{center}

\vskip .026in
\begin{center}
\r 1  {\eightit Institute of Physics, Academia Sinica, Taipei, Taiwan 11529, 
Republic of China} \\
\r 2  {\eightit Argonne National Laboratory, Argonne, Illinois 60439} \\
\r 3  {\eightit Istituto Nazionale di Fisica Nucleare, University of Bologna,
I-40127 Bologna, Italy} \\
\r 4  {\eightit Brandeis University, Waltham, Massachusetts 02254} \\
\r 5  {\eightit University of California at Los Angeles, Los 
Angeles, California  90024} \\  
\r 6  {\eightit Instituto de Fisica de Cantabria, University of Cantabria, 
39005 Santander, Spain} \\
\r 7  {\eightit Enrico Fermi Institute, University of Chicago, Chicago, 
Illinois 60637} \\
\r 8  {\eightit Joint Institute for Nuclear Research, RU-141980 Dubna, Russia}
\\
\r 9  {\eightit Duke University, Durham, North Carolina  27708} \\
\r {10}  {\eightit Fermi National Accelerator Laboratory, Batavia, Illinois 
60510} \\
\r {11} {\eightit University of Florida, Gainesville, Florida  32611} \\
\r {12} {\eightit Laboratori Nazionali di Frascati, Istituto Nazionale di Fisica
               Nucleare, I-00044 Frascati, Italy} \\
\r {13} {\eightit University of Geneva, CH-1211 Geneva 4, Switzerland} \\
\r {14} {\eightit Harvard University, Cambridge, Massachusetts 02138} \\
\r {15} {\eightit Hiroshima University, Higashi-Hiroshima 724, Japan} \\
\r {16} {\eightit University of Illinois, Urbana, Illinois 61801} \\
\r {17} {\eightit The Johns Hopkins University, Baltimore, Maryland 21218} \\
\r {18} {\eightit Institut f\"{u}r Experimentelle Kernphysik, 
Universit\"{a}t Karlsruhe, 76128 Karlsruhe, Germany} \\
\r {19} {\eightit Korean Hadron Collider Laboratory: Kyungpook National
University, Taegu 702-701; Seoul National University, Seoul 151-742; and
SungKyunKwan University, Suwon 440-746; Korea} \\
\r {20} {\eightit High Energy Accelerator Research Organization (KEK), Tsukuba, 
Ibaraki 305, Japan} \\
\r {21} {\eightit Ernest Orlando Lawrence Berkeley National Laboratory, 
Berkeley, California 94720} \\
\r {22} {\eightit Massachusetts Institute of Technology, Cambridge,
Massachusetts  02139} \\   
\r {23} {\eightit Institute of Particle Physics: McGill University, Montreal 
H3A 2T8; and University of Toronto, Toronto M5S 1A7; Canada} \\
\r {24} {\eightit University of Michigan, Ann Arbor, Michigan 48109} \\
\r {25} {\eightit Michigan State University, East Lansing, Michigan  48824} \\
\r {26} {\eightit University of New Mexico, Albuquerque, New Mexico 87131} \\
\r {27} {\eightit The Ohio State University, Columbus, Ohio  43210} \\
\r {28} {\eightit Osaka City University, Osaka 588, Japan} \\
\r {29} {\eightit Universita di Padova, Istituto Nazionale di Fisica 
          Nucleare, Sezione di Padova, I-35131 Padova, Italy} \\
\r {30} {\eightit University of Pennsylvania, Philadelphia, 
        Pennsylvania 19104} \\   
\r {31} {\eightit Istituto Nazionale di Fisica Nucleare, University and Scuola
               Normale Superiore of Pisa, I-56100 Pisa, Italy} \\
\r {32} {\eightit University of Pittsburgh, Pittsburgh, Pennsylvania 15260} \\
\r {33} {\eightit Purdue University, West Lafayette, Indiana 47907} \\
\r {34} {\eightit University of Rochester, Rochester, New York 14627} \\
\r {35} {\eightit Rockefeller University, New York, New York 10021} \\
\r {36} {\eightit Rutgers University, Piscataway, New Jersey 08855} \\
\r {37} {\eightit Texas A\&M University, College Station, Texas 77843} \\
\r {38} {\eightit Texas Tech University, Lubbock, Texas 79409} \\
\r {39} {\eightit Istituto Nazionale di Fisica Nucleare, University of Trieste/
Udine, Italy} \\
\r {40} {\eightit University of Tsukuba, Tsukuba, Ibaraki 305, Japan} \\
\r {41} {\eightit Tufts University, Medford, Massachusetts 02155} \\
\r {42} {\eightit Waseda University, Tokyo 169, Japan} \\
\r {43} {\eightit University of Wisconsin, Madison, Wisconsin 53706} \\
\r {44} {\eightit Yale University, New Haven, Connecticut 06520} \\
\end{center}

}
\draft
\address{}
\date{\today}
\maketitle
%
%

\begin{abstract}
We have studied the production of $B$ hadrons in 1.8-TeV \pbarp\
collisions. We present measurements of the fragmentation fractions,
$f_u$, $f_d$, $f_s$ and $f_{baryon}$, of produced $b$~quarks that
yield $B^+$, $B^0$, $B^0_s$ and \Lambdab\ hadrons.  Reconstruction of
several electron-charm final states yields
$f_s/(f_u+f_d)=0.213\pm0.068$ and
$f_{baryon}/(f_u+f_d)=0.118\pm0.042$, assuming $f_u=f_d$.  If all $B$
hadrons produced in \pbarp\ collisions cascade to one of these four
hadrons, we determine $f_u=f_d=0.375\pm0.023$, $f_s=0.160\pm0.044$ and
$f_{baryon}=0.090\pm0.029$.  If we do not assume $f_u = f_d$, we find
$f_d/f_u=0.84\pm0.16$.
\end{abstract}

%
%
\pacs{PACS Numbers: 13.60.Le, 13.60.Rj, 14.65.Fy, 13.87.Fh}

%
%
Bottom ($b$) quarks are not observed as independent entities but are
confined with a partner antiquark or diquark inside hadrons.  Once a
$b$ quark is produced, the process by which it combines with quarks
and gluons to form a hadron is called fragmentation and is governed by
the strong force, described by the theory of Quantum Chromodynamics
(QCD)~\cite{qcd}.  In this fragmentation process, the color force
field creates additional quark-antiquark partners that then combine
with the bottom quark to create a $B$~hadron.

The process by which a bottom quark fragments into a hadron cannot be
reliably calculated using perturbative QCD methods.  Therefore, the
fragmentation properties of the $b$ quark must be determined
empirically.  In this Letter, we investigate one such property, namely
the flavor dependence of the fragmentation process for bottom quarks
produced in 1.8-TeV $p\overline{p}$ collisions.  Our results provide
the most accurate measurements of this flavor dependence and for the
first time bring together in one study all previously studied $B$
hadrons.

We define $f_u$, $f_d$, $f_s$ and $f_{baryon}$ to be the probabilities
that the fragmentation of a $\overline{b}$ quark will result in a
weakly decaying $B^+$, $B^0$, $B^0_s$ meson and
$\overline{\Lambda}^0_b$ baryon, respectively.  We explicitly include
in these ``fragmentation fractions'' contributions from production of
heavier $B$ hadrons that decay into final states containing a $B^+$,
$B^0$, $B^0_s$ meson or $\overline{\Lambda}^0_b$ baryon.  The ALEPH
experiment used reconstructed $B_s^0\rightarrow D_s^- l^+ \nu X$
decays produced in $e^+e^-$ collisions at the $Z^0$ resonance to
determine the value $f_s = 0.120^{+0.045}_{-0.034}$~\cite{alephfs,pdg}.
The LEP Working Group on $B$ Oscillations has compiled $B^0
\overline{B}^0$ mixing results from the four LEP experiments and the
Collider Detector at Fermilab (CDF) experiment for the mixing
parameters $\overline{\chi}$ and $\Delta m_d$~\cite{pdg}.  The average
values of these parameters constrain the value of $f_s$, yielding the
result $f_s = 0.101^{+0.020}_{-0.019}$~\cite{pdg}.  The CDF experiment
has measured $f_s/(f_u+f_d)=0.210\pm0.036^{+0.038}_{-0.030}$ using
double semileptonic decays $b\rightarrow c\mu X$ with $c\rightarrow
s\mu X$~\cite{szymon}.  The ALEPH and DELPHI experiments measured
$f_{baryon}$ by reconstructing $\overline{\Lambda}_b^0\rightarrow
\Lambda_c^- l^+ \nu X$ decays~\cite{alephfb,delphifb}.  Their combined
result is $f_{baryon}=0.101^{+0.039}_{-0.031}$~\cite{pdg}.  The CLEO
experiment determined the quantity analogous to $f_d/f_u$,
$f^0/f^+=(\Upsilon(4S)\rightarrow
B^0\overline{B}^0)/(\Upsilon(4S)\rightarrow B^+B^-) = 0.88\pm0.16$ and
$0.90\pm0.14$, by reconstructing $B\rightarrow D^* l \nu$
decays~\cite{cleofufdsemi} and $B\rightarrow J/\psi K^{(*)}$
decays~\cite{cleofufd}, respectively.  Both of these measurements are
consistent with the isospin symmetry expectation that $f_d=f_u$.

Our measurement is performed by reconstructing $B$ hadron semileptonic
decays to electrons and charm hadrons from a $107$~pb$^{-1}$ sample of
1.8-TeV \pbarp\ collisions recorded by CDF during 1992-95.  The ratios
of the $b$ quark fragmentation fractions, namely $f_d/f_u$,
$f_s/(f_u+f_d)$ and $f_{baryon}/(f_u+f_d)$, are determined from the
$B$ hadron production ratios.  We reconstruct the $B$ hadrons in the
following decay modes and their charge conjugates: $B^+ \rightarrow
\overline{D}^0 e^+\nu_e X$ where $\overline{D}^0 \rightarrow K^+ \pi^-$;
$B^0 \rightarrow D^{*-} e^+\nu_e X$ where $D^{*-}\rightarrow
\overline{D}^0 \pi^-$ and $\overline{D}^0 \rightarrow K^+ \pi^-$; $B^0
\rightarrow D^- e^+\nu_e X$ where $D^-\rightarrow K^+ \pi^- \pi^-$;
$B^0_s \rightarrow D^-_s e^+\nu_e X$ where $D^-_s\rightarrow \phi \pi^-$
and $\phi\rightarrow K^+ K^-$; and $\overline{\Lambda}^0_b \rightarrow
\Lambda_c^- e^+\nu_e X$ where $\Lambda_c^-\rightarrow\overline{p} K^+
\pi^-$.
The average transverse momentum of the $B$ hadrons we reconstruct is
20~GeV/$c$.

The CDF detector has been described in detail elsewhere~\cite{ref: CDF
Detector}.  The CDF coordinate system defines the $z$ axis along the
proton beam direction and the polar angle $\theta$ with respect to the
$z$ axis.  The azimuthal angle $\phi$ is measured in the plane
transverse to the beam.  The relevant detector components for this
measurement are the charged-particle tracking system and the
calorimeters.  The tracking detectors lie inside a 1.4-T solenoidal
magnetic field.  The silicon vertex detector (SVX), positioned
immediately outside the beampipe, provides precise charged particle
reconstruction and allows identification of displaced vertices from
secondary decays.  The central tracking chamber (CTC), which
encompasses the SVX, measures the momenta of charged particles over a
pseudorapidity range $|\eta|<1.1$, where $\eta \equiv -\ln
\tan(\theta/2)$.  The central electromagnetic (CEM) and hadronic (CHA)
calorimeters, arranged in a projective tower geometry, surround the
tracking volume and are used to measure clusters of energy deposited
by electrons, photons and hadrons.  The central electromagnetic strip
chamber (CES), embedded in the CEM at the position of shower maximum,
measures the electromagnetic shower profiles in the $\phi$ and $z$
directions.

A three level trigger system is used to identify electron candidates,
with the first level requiring a CEM energy deposition greater than 8
GeV.  The electron candidates satisfy a Level~2 trigger that requires
a spatial match between a track in the CTC with $P_T>7.5$~GeV/{\it c}
and an energy cluster in the CEM with $E_T>8.0$~GeV, where $P_T\equiv
P\sin\theta$ and $E_T\equiv E\sin\theta$.  The fraction of hadronic
energy in the cluster is required to be small.  We require a spatial
match of the CTC track to a cluster of energy in the CES and apply a
threshold requirement to the energy deposition in the CES.  The
Level~3 trigger requires that the lateral shower profiles in the CES
and CEM be consistent with those expected for an electron, and
re-applies the previous trigger criteria with greater precision.
Approximately six million electron candidates survive this trigger
selection.  We reduce the sample to three~million candidates by
applying more stringent criteria~\cite{thesis}.  We require that the
fraction of hadronic energy in the cluster be less than 4\%.  We
reject electron candidates that are likely to be from photon
conversions and from $W^\pm$ and $Z^0$ boson decays.  Finally, to
ensure a uniform electron identification efficiency in the different
$B$ hadron decay topologies, we reject candidates with more than one
track pointing at the CEM cluster and demand that the ratio of cluster
energy to track momentum be in the range $0.75<E_T/P_T<1.40$.

The semileptonic $B$ hadron decays are identified by reconstructing
the charm hadron in the vicinity of the electron.  The
$\overline{D}^0$ meson is reconstructed by identifying the products of
the $\overline{D}^0\rightarrow K^+\pi^-$ decay in a cone
$R\equiv\sqrt{(\Delta\eta)^2+(\Delta\phi)^2} = 1.0$ around the
electron track.  The charge correlation between the electron and the
charm hadron daughters from semileptonic $B$ hadron decays is
exploited to reduce the contamination from random combinations of
leptons and charmed hadrons.  Particles arising from the weak decay of
a $B$~hadron are normally displaced from the primary vertex.
Therefore, we require the charm-hadron daughter tracks to have an
impact parameter ($d_0$) inconsistent with zero
($|d_0|/\sigma(d_0)>1.5$, where $\sigma(d_0)$ is the uncertainty on
$d_0$).  The combinatorial background is further reduced by requiring
that $P_T(K)>1.2$~GeV/{\it c} and $P_T(\pi)>0.5$~GeV/{\it c}, which
are the same criteria used in the reconstruction of the other
channels, except where noted.  The daughter tracks are required to be
consistent with coming from a secondary vertex that is displaced in
the transverse plane from the \pbarp\ interaction point
($L_{xy}/\sigma(L_{xy})>1$).  Finally, the invariant mass of the
electron-charm system is required to be less than 5.0~GeV/$c^2$.

\begin{figure}
\vspace*{5.0in}
\hspace*{-0.5in}
\vbox{
\includegraphics{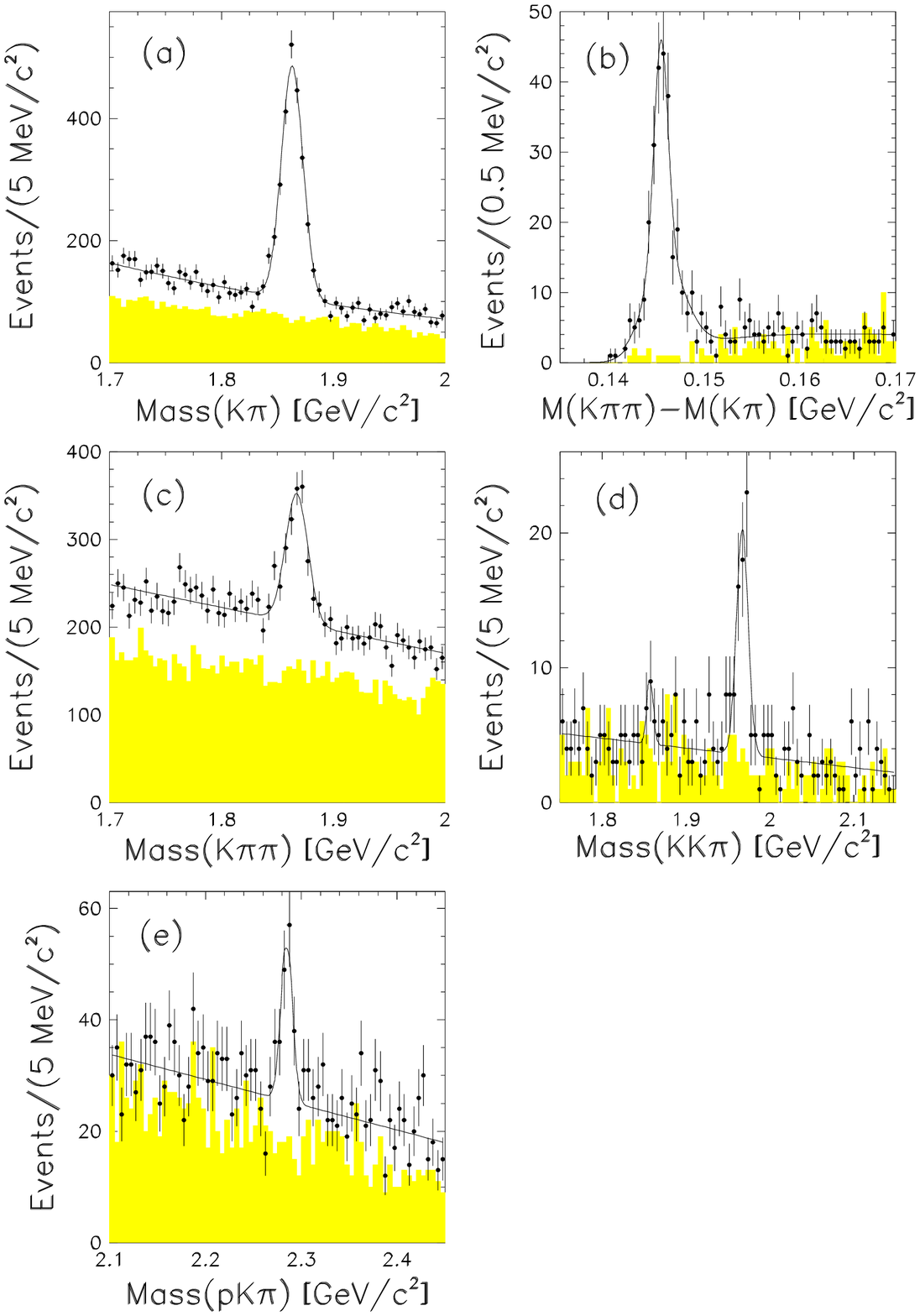}
}
\caption
{Invariant mass distributions of charm hadron candidates in
107~pb$^{-1}$ of inclusive electron data.  a) $K\pi$ invariant mass
distribution for $\overline{D}^0$ candidates.  b) Mass difference
distribution, $\Delta M = M(K\pi\pi)-M(K\pi)$, for $D^{*-}$
candidates. c) $K\pi\pi$ invariant mass distribution for $D^-$
candidates. d) $KK\pi$ invariant mass distribution for $D_s^-$
candidates. e) $pK\pi$ invariant mass distribution for $\Lambda_c^-$
candidates.  The shaded histograms represent the combinations with the
wrong electron-hadron charge correlation.  The shaded area in a) has
been scaled by 0.5 for display purposes.}
\label{fig:mass}
\end{figure}

The invariant mass of the $K\pi$ candidates in the electron sample is
shown in Fig.~\ref{fig:mass}(a).  To this distribution we fit the sum
of a Gaussian signal and an exponential background and count
$1848\pm58$ $\overline{D}^0$ signal events.  We observe no significant
signal in the combinations with the wrong electron-hadron charge
correlation.

The $D^{*-}$ meson is reconstructed in the $D^{*-}
\rightarrow\overline{D}^0\pi^-$ channel. We consider $\overline{D}^0$
candidates with $1.80< M(K\pi)<1.95$~GeV/{\it c}$^2$, where $M$ is the
mass, and consider all charged particles with $P_T>0.4$~GeV/{\it c}\
for the additional pion.  The mass difference distribution, $\Delta M
= M(K\pi\pi)-M(K\pi)$, is shown in Fig.~\ref{fig:mass}(b).  To this
distribution we fit a double-Gaussian signal and a background modeled
by a threshold function.  We reconstruct $249\pm19$ $D^{*-}$ signal
events.

The $D^-$ meson is reconstructed in the $D^-\rightarrow K^+\pi^-\pi^-$
channel.  In this channel, the three daughter tracks are required to
form a vertex.  The invariant mass distribution of the $K\pi\pi$
candidates is shown in Fig.~\ref{fig:mass}(c).  To this distribution
we fit the sum of a Gaussian signal and a linear background and count
$736\pm62$ $D^-$ signal events.

The $D_s^-$ meson candidates are identified by looking for the
products of the $D_s^-\rightarrow \phi \pi^-$ decay, where $\phi
\rightarrow K^+K^-$.  Both kaons and the pion are required to come
from a common vertex.  This decay chain provides two additional
criteria effective in rejecting combinatorial backgrounds.  First, we
require that the mass of the $K^+K^-$ system be within 0.010~GeV/{\it
c}$^2$ of the world average $\phi$ mass of 1.019~GeV/{\it c}$^2$.
Second, we impose the criterion $|\cos\psi|>0.4$, where $\psi$ is the
helicity angle between the $D_s$ and $K$ meson candidates in the $\phi$ rest
frame.  The invariant mass distribution of the $KK\pi$ candidates is
shown in Fig.~\ref{fig:mass}(d).  We reconstruct $59\pm10$ $D_s^-$
signal events.  

The $\Lambda_c^-$ baryon candidates are identified by looking for the
products of the $\Lambda_c^-\rightarrow \overline{p} K^+ \pi^-$ decay.
We require $P_T(p)>2.0$~GeV/{\it c}.  Since the relative combinatorial
background under the $\Lambda_c^-$ signal is large, we also require
that the specific ionization ($dE/dx$) deposited by the proton
candidate in the CTC be consistent with that expected for a 
proton.  The invariant mass distribution of the $pK\pi$ candidates is
shown in Fig.~\ref{fig:mass}(e).  We reconstruct $79\pm17$
$\Lambda_c^-$ signal events.

The $D_s^- e^+$ and $\Lambda_c^- e^+$ final states represent
relatively pure samples of the $B^0_s$ and $\overline{\Lambda}_b^0$
hadrons, respectively.  However, the remaining electron-charm
final states that we reconstruct come from several $B$-meson species
through feed-down from vector and orbitally-excited $D$ meson decays.
For example, the decay $B_s^0\rightarrow D_s^{**-}e^+\nu_e $ can be
followed by the decay $D_s^{**-}\rightarrow D^-K^0$.  This channel
contributes to the $D^-e^+$ sample but reflects $B_s^0$ production
rather than $B^0$ production.  We use the branching fractions for each
decay to determine the feed-down contributions.  These branching
fractions are derived from the measured values~\cite{pdg} according to
the spectator model and isospin symmetries~\cite{thesis}.

The spectator model predicts that the inclusive semileptonic decay
widths for the various $B$ hadrons are equal, yielding, for example,
the relation
\begin{displaymath}
{\cal B}(B_s^0 \rightarrow e^+ \nu_e X) =
\frac{\tau(B^0_s)}{\langle\tau(B)\rangle}{\cal B}(B \rightarrow e^+ \nu_e X),
\end{displaymath}
where ${\cal B}$ represents the branching fraction and $\tau$ is the
lifetime.  A similar relationship holds for the exclusive semileptonic
branching fractions for the three $B$ mesons.  We use isospin symmetry
to calculate the branching fractions for the $D^*$ and $D^{**}$ decays
that feed down into the final state $D$ mesons that we reconstruct.

The acceptance and reconstruction efficiency for each final state vary
according to whether the $D$ meson is produced directly in the $B$
meson decay or is the daughter of an excited $D$ meson state.  Several
efficiencies, such as the electron identification efficiency, the
conversion removal efficiency and the two-track finding efficiency,
cancel in the ratio of fragmentation fractions.  Of the remaining
efficiencies, the single-track finding efficiency and the electron
trigger efficiency are measured using CDF data.  We use a Monte Carlo
calculation to determine all other acceptances and efficiencies.  This
uses a next-to-leading order perturbative calculation of the
differential cross section for $b$-quark production in \pbarp\
collisions~\cite{NDE} followed by fragmentation governed by the
Peterson formulation~\cite{peterson}.  We use the ISGW~\cite{isgw}
model to determine the semileptonic $B$ hadron decay
kinematics~\cite{cleomc}.

The systematic uncertainties on the reconstruction efficiencies
include those associated with the tracking and trigger efficiencies
and Monte Carlo statistics.  We also assign an uncertainty to account
for the poor knowledge of the $\overline{\Lambda}_b^0$ production
polarization in \pbarp\ collisions.  Finally, we consider the
possibility that the Peterson fragmentation parameter $\epsilon_b$ may
differ for each $B$ hadron species.  We assign the fractional
uncertainties of 6.4\% and 6.1\% to account for the possible variation
of $\epsilon_b$ for $B_s^0$ and $\overline{\Lambda}_b^0$ hadrons,
respectively.  We determine these values by calculating the larger
variation in the reconstruction efficiency for values of $\epsilon_b$
one standard deviation away from a central value of
$\epsilon_b=0.006\pm0.002$~\cite{chrin}.  These contributions
represent the largest uncertainties associated with the reconstruction
efficiencies.  The total fractional systematic uncertainties are
$\pm2.1$\%, $\pm3.4$\%, $\pm5.4$\%, $\pm7.3$\%, and $\pm8.7$\% for
$B^+ \rightarrow \overline{D}^0 e^+\nu_e X$, $B^0 \rightarrow D^- e^+\nu_e
X$, $B^0 \rightarrow D^{*-} e^+\nu_e X$, $B^0_s \rightarrow D^-_s e^+\nu_e
X$ and $\overline{\Lambda}^0_b \rightarrow \Lambda_c^- e^+\nu_e X$
decays, respectively.

In order to determine the fragmentation fractions taking into account
the cross contamination and feed-down, we fit the five observed event
yields and their uncertainties to the three ratios of fragmentation
fractions.  We formulate the problem by defining a $\chi^2$ function
comparing the predicted with observed event yields.  We allow the
semileptonic branching fractions for the $B$ mesons to vary in the
fit, constrained to their measured or calculated uncertainties.  We
note that the measured branching fractions include the implicit
assumption that $f^0/f^+=1$.

We make this measurement assuming isospin symmetry by fixing $f_d/f_u
= 1$ in the fit.  The fit results in the values $f_s/(f_u+f_d) =
0.213\pm0.068$ and $f_{baryon}/(f_u+f_d) = 0.118\pm0.042$, where
uncertainties on the event yields, the reconstruction efficiencies and
the branching fractions are included.  We can determine the absolute
fragmentation fraction values from our fits by assuming that the
$B^0$, $B^+$, $B_s^0$ and $\overline{\Lambda}_b^0$ hadrons saturate
the $b$-quark production rate, i.e., $f_u + f_d + f_s + f_{baryon}
\equiv 1$.  This relationship yields $f_u = f_d = 0.375 \pm 0.023$,
$f_s = 0.160 \pm 0.044$ and $f_{baryon} = 0.090 \pm 0.029$.  By
incorporating another term in the $\chi^2$ function, we have combined
these results together with the complementary measurement by CDF of
$f_s$ using double semimuonic decays~\cite{szymon} to determine the
more precise values of $f_u=f_d=0.375\pm0.015$, $f_s=0.160\pm0.025$
and $f_{baryon}=0.090\pm0.028$.  Results for the fragmentation
fractions obtained using all available measured exclusive semileptonic
branching fractions instead of employing the spectator model
predictions are consistent with the results presented here.

By relaxing the isospin symmetry requirement, we find that $f_d/f_u =
0.84\pm0.16$, consistent with isospin symmetry and the measurements of
$f^0/f^+$ by the CLEO collaboration.  The individual values for $f_s$
and $f_{baryon}$ remain unchanged.

In conclusion, we have measured the four $b$ quark fragmentation
fractions for weakly decaying $B$ hadrons produced in \pbarp\
collisions.  We measure $f_{baryon}=0.090\pm0.029$, in good agreement
with measurements made on $B$~hadrons produced in high energy
$e^+e^-$\ collisions at LEP. The \pbarp\ result, $f_s =
0.160\pm0.025$, is two standard deviations higher than the current
world average value, which is dominated by LEP measurements and by
inference from $B^0\overline{B}^0$ mixing measurements.

We thank the Fermilab staff and the technical staff at the
participating institutions for their essential contributions to this
research.  This work is supported by the U.~S.~Department of Energy
and the National Science Foundation; the Natural Sciences and
Engineering Research Council of Canada; the Istituto Nazionale di
Fisica Nucleare of Italy; the Ministry of Education, Science and
Culture of Japan; the National Science Council of the Republic of
China; and the A.~P.~Sloan Foundation.

%
%



%
%
%
%
\end{document}